\documentclass{aa}  
\usepackage{graphicx}

\usepackage{txfonts}

\def\0{\phantom0}

\def\zl{$z_{\rm lens}$}
\def\zqso{$z_{\rm qso}$}
\def\ho{H$_{\rm 0}$}

\def\kmsmpc{km s$^{-1}$ Mpc$^{-1}$}
\def\om{$\Omega_{\rm{m}}$}
\def\ol{$\Omega_{\rm{\Lambda}}$}
\def\obj{SDSS J1650+4251}

\begin{document}

   \title{COSMOGRAIL: the COSmological  MOnitoring of \\ \vspace*{1mm}
   GRAvItational Lenses}
   
   \subtitle{V. The time delay in \obj}
   
   \titlerunning{COSMOGRAIL~V: time delay measurement in \obj}

   \author{C. Vuissoz\inst{1} \and F. Courbin\inst{1} \and D. Sluse\inst{1} 
        \and G. Meylan\inst{1}  \and M. Ibrahimov\inst{2} \and
        I. Asfandiyarov\inst{2} \and E. Stoops\inst{1,3} \and A. Eigenbrod\inst{1}
       	\and L. Le Guillou\inst{4} \and H. van Winckel \inst{4}
	\and P. Magain\inst{3}}


    \institute{
     Laboratoire d'Astrophysique, Ecole Polytechnique F\'ed\'erale
     de Lausanne (EPFL), Observatoire, CH-1290 Sauverny, Switzerland
     \and
     Ulugh Beg Astronomical Institute, Academy of Sciences, Tashkent, Uzbekistan
     \and 
     Institut d'Astrophysique et de G\'eophysique, Universit\'e de
     Li\`ege, All\'ee du 6 ao\^ut 17, Sart-Tilman, Bat B5C, B-4000 Li\`ege,
     Belgium
     \and
     Instituut voor Sterrenkunde, Katholieke Universiteit Leuven, 
     Celestijnenlaan 200B, B-3001 Heverlee, Belgium}

   \date{Received ... ; accepted ...}

 
  \abstract 
 {}
    {Our aim is to measure the time delay between the two
      gravitationally lensed images of the \zqso\ = 1.547 quasar \obj,
      in order to estimate the Hubble constant \ho.}
    {Our  measurement  is  based on $R$-band light curves with 57
      epochs obtained at Maidanak Observatory,  in  Uzbekistan, from
      May 2004 to September  2005.   The  photometry is    performed
      using   simultaneous deconvolution  of the data,  which provides
      the individual light curves of the  otherwise blended quasar
      images. The  time delay is determined from the light curves
      using two very different numerical techniques, i.e., polynomial
      fitting and direct cross-correlation. The time delay is
      converted into \ho\ following analytical modeling of the
      potential well.}
    {Our best  estimate  of the time  delay  is $\Delta t  = 49.5 \pm
      1.9$~days, i.e., we reach a 3.8\,\% accuracy. The $R$-band flux
      ratio between the quasar images, corrected for the time delay
      and for slow microlensing, is $F_{\rm A}/F_{\rm  B} = 6.2
        \pm 5$\,\%.}
{The  accuracy reached on the  time delay allows us to discriminate well
between  families of  lens models.  As  for most other multiply imaged
quasars, only models  of the   lensing galaxy  that  have a de
Vaucouleurs mass profile plus external shear give a Hubble constant
compatible with the current most popular value (\ho\  = 72 $\pm$ 8 \kmsmpc).
A more realistic singular isothermal sphere model plus external shear
gives \ho\ = 51.7$^{+4.0}_{-3.0}$ \kmsmpc.}


\maketitle

\section{Introduction}

\begin{figure*}[ht]
\begin{center}
\includegraphics[width=17.8cm]{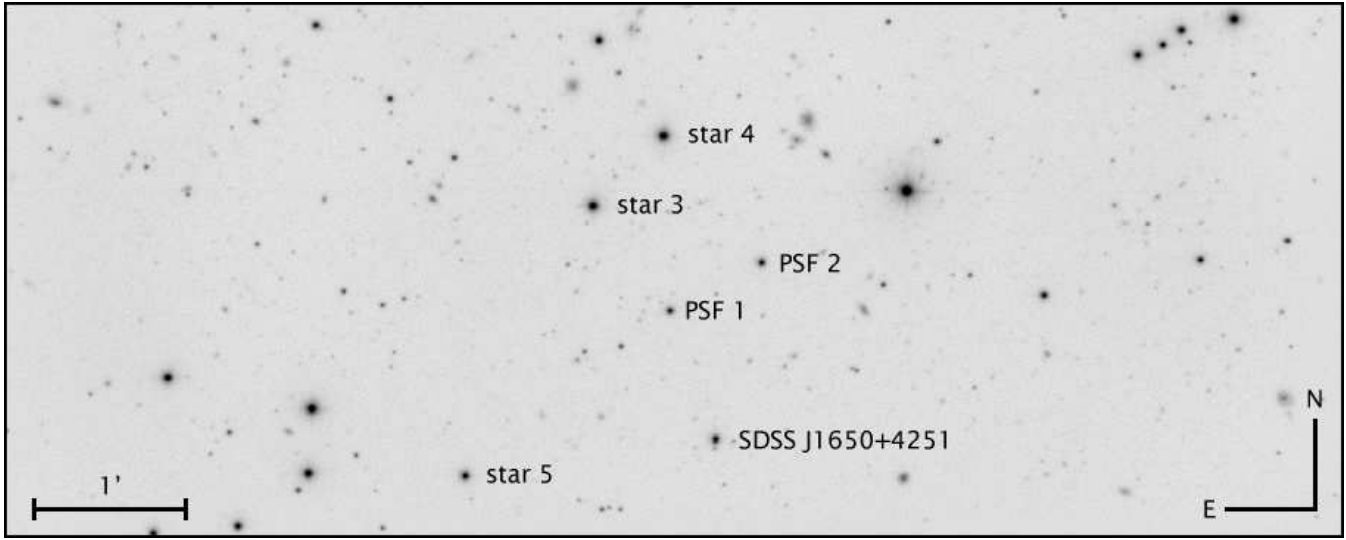}
\caption{A $R$-band image of \obj\ obtained at Maidanak Observatory. 
  This image  is a combination of 376  frames, totalising  31 hours of
  exposure.  The mean seeing  is 1.0\arcsec, and the  field of view is
  $3.5\arcmin \times   8.9\arcmin$.  The two  stars  labeled PSF1 and
  PSF2 are used  to model the Point Spread  Function required for  the
  MCS  deconvolution method.  The 4  reference stars used for the photometric
  calibration are star \#3, \#4, PSF1 and PSF2.  Star \#5
  is used as a cross-check of the deconvolution photometry.}
\label{field}
\end{center}
\end{figure*}

The so-called  time-delay method in  gravitationally  lensed quasars
(Refsdal \cite{Refsdal64})  is one  of the  rare techniques   that can
yield a measurement of the Hubble constant  \ho\ at truly cosmological
distances, independently of any local calibration or standard  candle.
Until now, however, no  concerted and long-term project has  succeeded
in applying it in a systematic way, at a level really competitive with
other techniques, such as Cepheids, supernovae  or the fluctuations of
the cosmic   microwave background radiation.  The   COSMOGRAIL project
aims  at measuring time  delays  for most known gravitationally lensed
quasars,  as well  as  improving  their lens  models, using  deep high
resolution imaging and spectroscopy.  The contribution of the error on
the time delay measurement to the total error budget  on \ho\ is about
50\,\%,  the  other half  being   due  to degeneracies   in  the  lens
models. The  immediate goal of COSMOGRAIL  is to make  the former
close to negligible in front  of the latter.  While  this is very hard
for lenses with short time delays (Kochanek et al.~\cite{Kochanek06a},
Morgan et al.~\cite{Morgan06}),  longer  time delays can  be  measured
accurately (Eigenbrod et al.~\cite{Eigenbrod05}).

In this paper, we present the first result from our monitoring program:
the  time delay   measurement in  the   gravitationally lensed  quasar
\obj\ (16$^\textrm{\scriptsize{h}}$ 50$^\textrm{\scriptsize{m}}$ 43\fs5,
 +42\degr\ 51\arcmin\ 45\farcs 0; J2000.0),  
discovered by Morgan et  al.  (\cite{Morgan03})  as a
doubly imaged \zqso\  = 1.547 quasar,   with an angular  separation of
1.2\arcsec.  At the time of the discovery,  the $B$-band magnitudes of
components  A and B were $17.8$  and $20.0$ respectively.  The lensing
galaxy  is  detected  by  Morgan  et   al.   (\cite{Morgan03})  in the
$I$ band. Absorption  lines seen in the spectrum  of the quasar images
suggest a lens redshift of \zl\ = 0.577. In a cosmology with \ho\ = 75
km s$^{-1}$ Mpc$^{-1}$,  Morgan et al.  (\cite{Morgan03}) predict that
the time delay between the  quasar images is  of the order of a month,
assuming a Singular Isothermal  Sphere (SIS) potential for the lensing
galaxy,  plus an external shear.   Following the non-parametric models
of Saha  et al.  (\cite{Saha06})  and the same cosmology, the expected
time delay  between  the two quasar   images is $\Delta  t \sim 30-60$
days.

The  first two  seasons of the  photometric  monitoring of  \obj\ were
carried  out with the  1.5-m   telescope at Maidanak  Observatory,  in
Uzbekistan.  Given  the  high declination of  \obj\   and the pointing
limits of the telescope, \obj\ can be followed for about 8 months per 
year under  good airmass and  seeing conditions.  This combination of 
time delay and visibility window makes
\obj\ an excellent target for an  accurate time delay measurement.  From
numerical simulations  using artificial light  curves, Eigenbrod et al.
(\cite{Eigenbrod05}) predict that the  accuracy on the time  delay can
be  as good as  1\,\%, assuming that  the peak to  peak amplitude of the
quasar light variation  is  0.2~mag over  the two  years  of observation and
using  a  temporal  sampling of  at   least  one observing  point  per
week. Although  more than two  observing seasons will  be necessary to
achieve this final  goal, it is  already possible to measure the  time
delay with 3.8\,\% accuracy, which is sufficient to show that lens models
 with constant mass-to-light ratio and models with dark matter profiles that
do not trace light give discrepant values of \ho.

The  photometric monitoring, the data reduction   and the light curves
are presented in Sects. 2 and 3 respectively. The determination of the time delay
between the two lensed images is described in Sect. 4.   In Sect.  5,
we discuss the mass models for the lensing galaxy and the implications
for   the  Hubble constant.   Finally, Sect.   6  summarises  the main
results.


\section{Observations}
\label{observations}

The  observations    presented in this  article    consist of two full
observing seasons.  \obj\  is visible from the end  of February to the
beginning   of   October at   Maidanak   Observatory.  The   resulting
non-visibility window is therefore  of about 150 days.  Additional, but
smaller,  gaps  are due  to   bad seeing  or  weather  conditions.  The
photometric points presented here span a total of 333 days between May
2004 and  September 2005, after removal  of the non-visibility period.
The mean temporal sampling in the visibility window is one point every
5th day.

The CCD camera used at Maidanak Observatory is a 800$\times$2000 array
with a   pixel size of 0.266\arcsec\   on the plane   of the sky.  The
useful field  of  view is 3.5\arcmin$\times$8.9\arcmin.  All  data are
taken through   the Cousins  $R$  filter.   The  seeing varies between
0.7\arcsec\  and   2.0\arcsec\   over  the   two    observing seasons,
1.1\arcsec\ being the most frequent value.  For each observing epoch,
unless technical or meteorological problem occurs, 6 dithered images are
taken.  The exposure time for each  of the 6 frames is  300 s, and the
size of the dithering box is 15\arcsec.


\section{Reductions and deconvolution}
\label{reduction}

\begin{figure*}[t]
\begin{center}
\includegraphics[width=15cm]{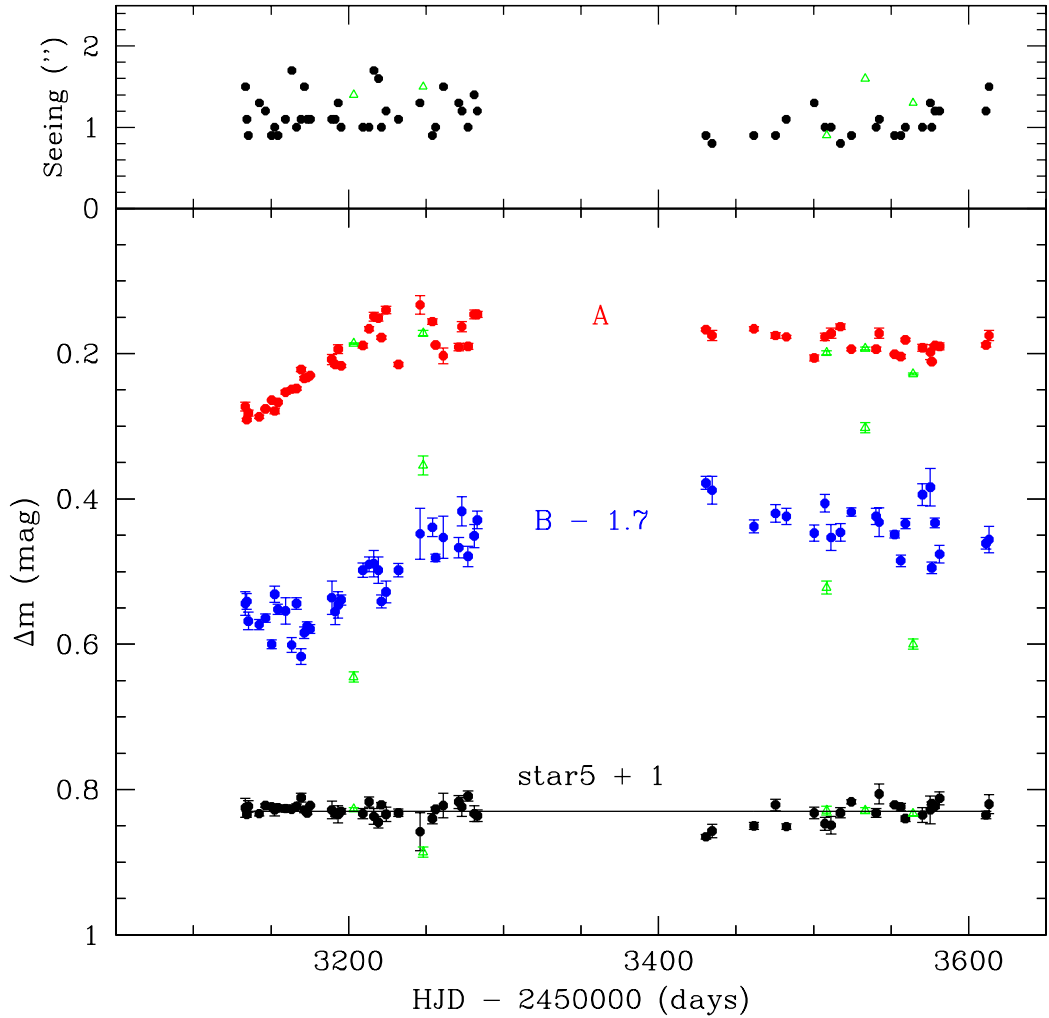}
\caption{$R$-band light curves for the two quasar images in \obj, as 
well as for a reference star in the field of view.  The magnitudes are
given in relative units.  In order to  avoid the points  overlaps, the
component B and the star curves have been shifted in magnitude.  The 5
epochs marked by triangles  deviate very far  away from  the otherwise
smooth variations of  the light curve  of component B. They  have been
removed  from the   curves    when determining the    time  delay (see
Section~\ref{timedelays}).}
\label{lightcurve}
\end{center}
\end{figure*}

The field of view of the camera is shown in Fig.~\ref{field}, where
\obj\ has been slightly off-centered in order to include more stars 
useful for the  construction of the Point  Spread  Function (PSF).  An
automated pipeline is  used to carry  out the pre-reduction of all the
individual CCD  frames.   This pipeline first  subtracts a master-bias
from each image, then for each night, corrects for the flatfield using
twilight flats, and removes a sky-image.  The accurate positions of 15
reference  stars are   then measured  on   all images  with  the  {\tt
Sextractor} package  (Bertin \& Arnouts  \cite{Bertin96}) and  used in
{\tt IRAF} \footnote{{\tt IRAF} is distributed by the National Optical
Astronomy  Observatories,  which are  operated  by the  Association of
Universities   for Research  in   Astronomy,  Inc.,  under cooperative
agreement with the   National  Science  Foundation.}  to  compute  the
geometrical transformation between the  images. Rebinning of the  data
using  polynomial  interpolations is    necessary, since  rotation and
scaling are  included in the transformation, in  addition to the image
shift.  We do not perform any cosmic ray  removal, since no cosmic ray
is present  in  the tiny field  of  view  considered around \obj\  and
around  the   PSF stars,   as indicated  by   the residual  images  of
the deconvolution.

Since the  airmass and the transparency  of the  sky change with time,
cross-calibration of the photometric zero point between all the images
is necessary prior to any magnitude measurement. We find that the best
way  to carry out  this  task on images  taken  at Maidanak, is to use
reference stars   as close as possible   to the  target.   This choice
leaves  us  with few calibration  stars  but the determination  of the
photometric offsets achieved in  this way remains  better than the one
using many more reference stars  located further away from the quasar.
We use four non-variable stars, labeled PSF 1, PSF 2, star \#3 and \#4
in Fig.~\ref{field}.

The  calibrated frames  are simultaneously  deconvolved using the  MCS
algorithm   (Magain   et  al.  \cite{Magain98}),  which   provides the
photometry of the two blended quasar images, free  of any mutual light
contamination.  This procedure  has already been  successfully applied
to the monitoring   data of  several  lensed  quasars  (Burud  et  al.
\cite{burud2000,   burud2002a,       burud2002b},       Hjorth      et
al. \cite{hjorth2002}, Jakobsson et al. \cite{Jakobsson05}).  Its main
advantage is  its ability to  deconvolve all the frames from different
epochs simultaneously,  constraining well  the  positions of  the  two
quasar  images even when  part of the data  is obtained in poor seeing
conditions.  In addition, no prior knowledge is  used on the shape and
position of the  lensing galaxy and  on the lensed  host galaxy of the
quasar.   All extended objects are treated  as a fully numerical array
of pixels.  The  flux of the quasar  images, treated analytically, are
allowed to vary from one frame to another,  hence leading to the light
curves.  
The resolution in the deconvolved image is a parameter given to the
algorithm: the deconvolution method splits each pixel of the initial
image in 4 parts of the same size, and the final resolution
corresponds to the size of two of these small pixels, accordingly to
the sampling theorem. Therefore it is given by the detector pixel
size, i.e. 0.27\arcsec.
We use two stars  to construct the  PSF required for the  MCS
deconvolution    to  work. They  are   labeled   PSF1    and PSF2   in
Fig.~\ref{field}.  They are also used as flux calibrators.

Although the two quasar images  are well separated in our  deconvolved
image with a final resolution of 0.27\arcsec, both the lensing galaxy 
and the lensed host galaxy of the quasar are too faint to be detected.

The $R$-band light curves obtained  using the deconvolution photometry
are presented  in Fig.~\ref{lightcurve} and  in Table~1.  They consist
of 62 data points.  Each point corresponds  to one given night, and is
the mean of 6 independent consecutive measurements.  The error bar for
each epoch  is the 1$\sigma$ standard error  on this mean value.  This
empirical  way    of   determining the  error   from   six independent
measurements obtained using deconvolution, ensures that PSF errors are
propagated in the final photometry in a realistic way.

We also show in Fig.~\ref{lightcurve} the photometry obtained via the
deconvolution of  the isolated star  \#5. Its light  curve is flat and
the standard  deviation between  all the epochs  ($\sigma_{tot}=0.006$
mag) is  compatible with  the mean  of the error  bar  on each
individual epoch ($\sigma_{mean}=0.007$ mag).


\section{Measurement of the time delay}
\label{timedelays}

\begin{figure}[t]
\begin{center}
\leavevmode
\includegraphics[width=4.3cm]{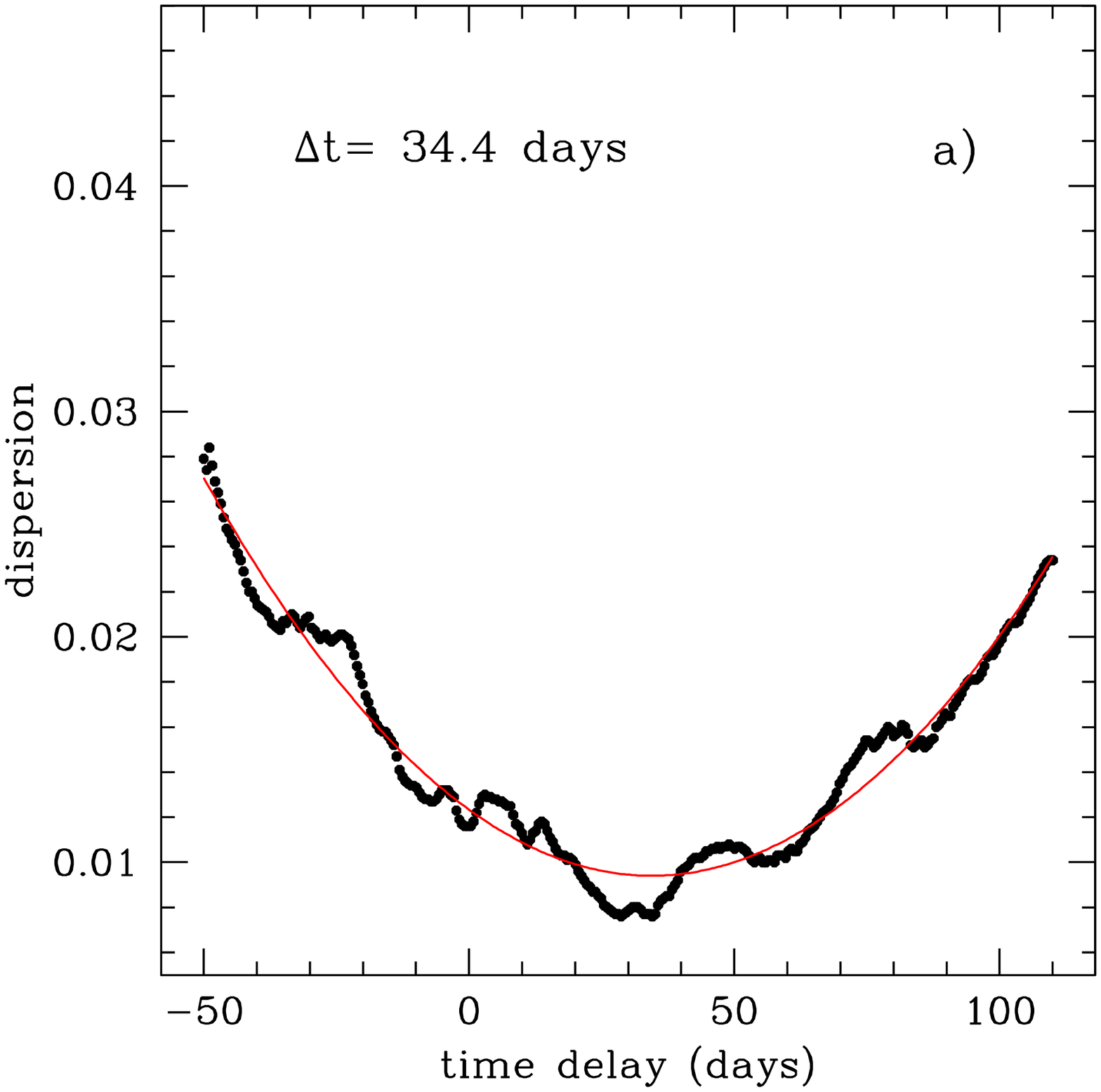}
\includegraphics[width=4.3cm]{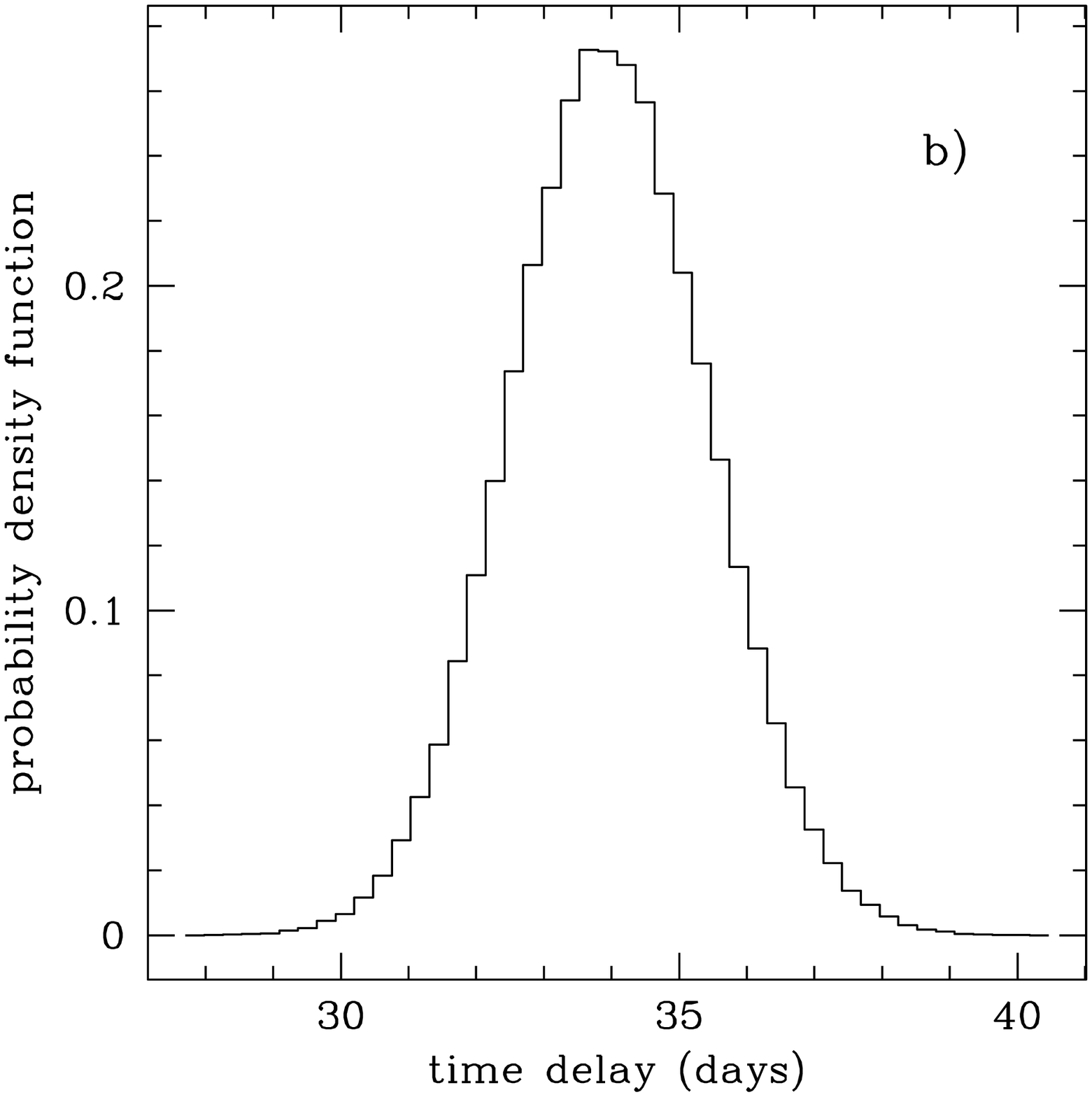}
\includegraphics[width=4.3cm]{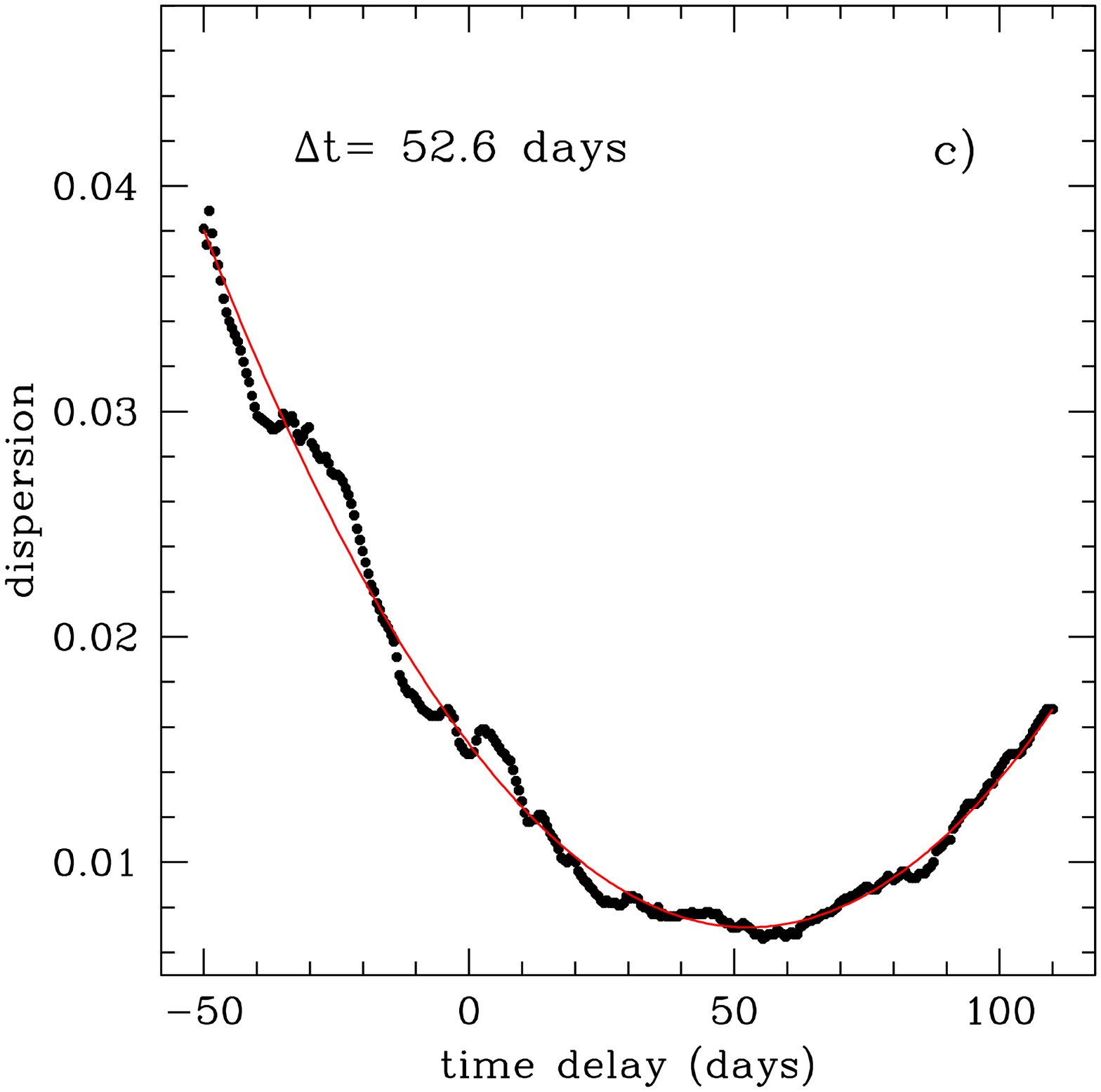}
\includegraphics[width=4.3cm]{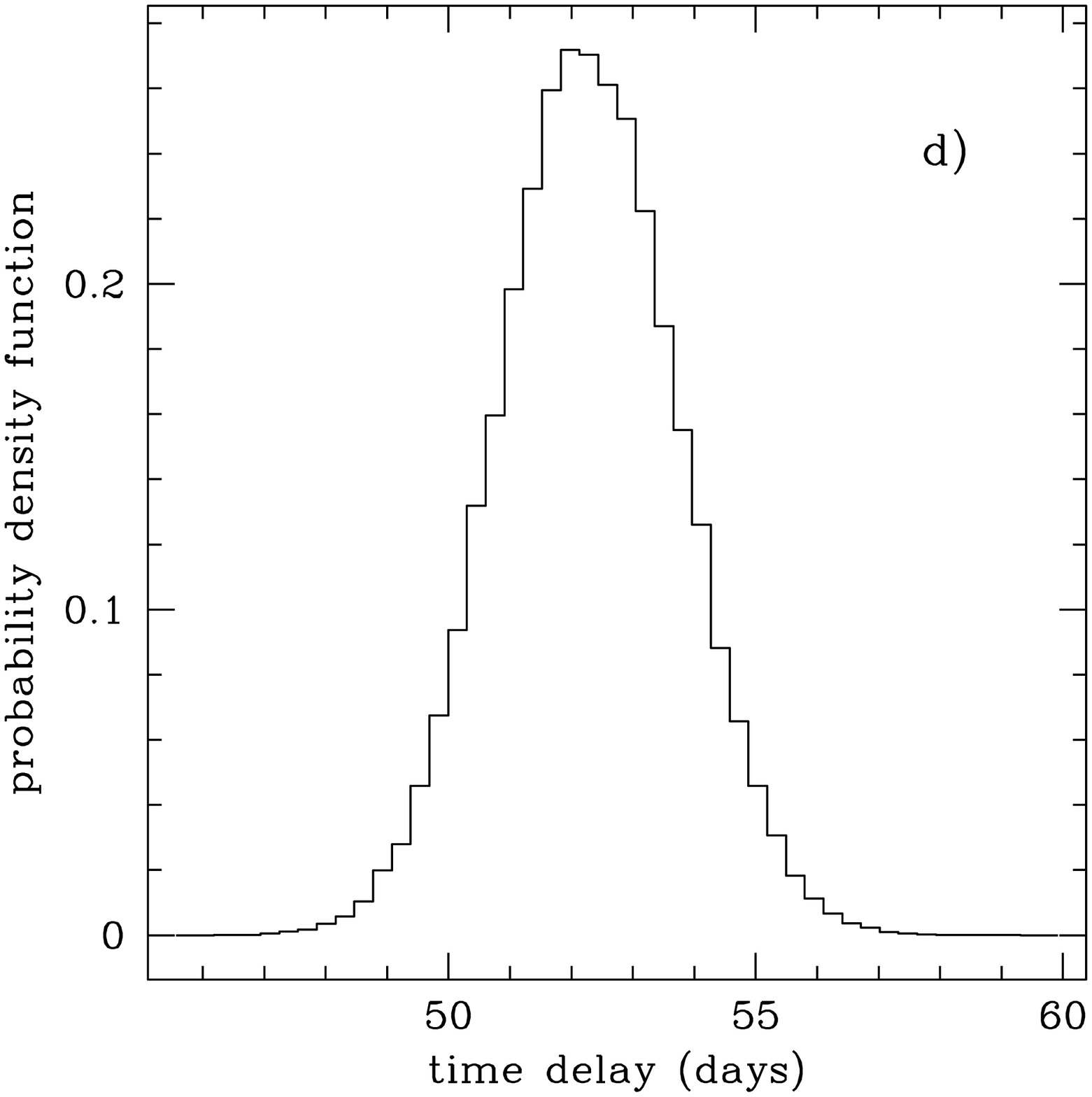}
\caption{a) The dispersion function obtained when the normalisation 
of the light curves is determined  directly from the data.  Several 
local minima appear. A polynomial fit  to the dispersion function 
(solid line) allows the determination  of a time  delay of 
$\Delta t \sim 34.4$~days. 
b) The corresponding result of the Monte-Carlo simulation for 100'000
slightly modified   light  curves:  $\Delta t \sim 34.4 \pm 1.4$~days. 
c) The dispersion function obtained for the light curves optimally shifted
in magnitude (see text and Fig.~\ref{TDresults2}).
The function is now much smoother and has only one clear minimum at
$\Delta t \sim 52.6$~days. 
d) The corresponding Monte-Carlo simulation result and the final 
time delay estimate: $\Delta t \sim 52.3 \pm 1.6$~days.
}
\label{TDresults1}
\end{center}
\end{figure}

\begin{figure}[t]
\begin{center}
\includegraphics[width=9.0cm]{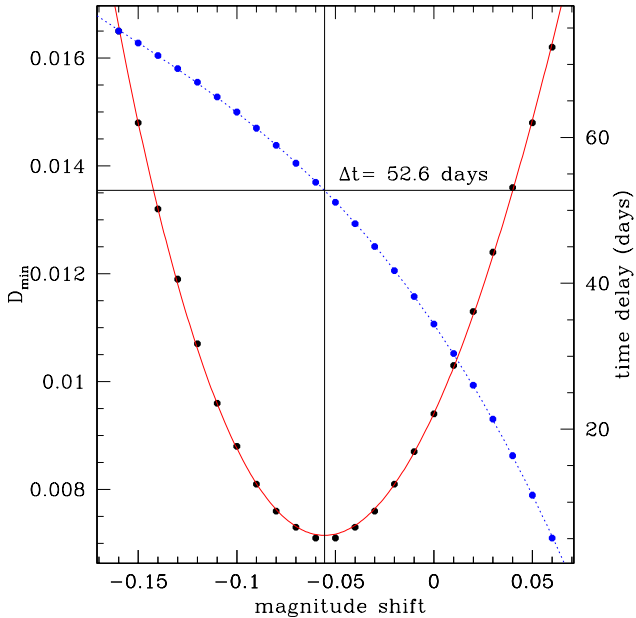}
\caption{Value of the minimum D$_{min}$ of the dispersion function, as 
a function of the magnitude shift between the two  light curves. A fit
(solid line) to the  measurements yields  the optimal magnitude  shift
$\Delta  m = -0.056$.   In the figure, a shift  of 0 corresponds to the
normalisation carried out  directly on the  data.  The dotted line and
the vertical  axis on the  right  give the  corresponding value of the
time delay $\Delta t \simeq  52.6$~days (for one single realisation of
the  Monte-Carlo simulation).  
}
\label{TDresults2}
\end{center}
\end{figure}

A  rough guess of the  light  curves shift indicates   a time delay of
about 50 days, i.e.,  20 days longer than  predicted in  the discovery
article (Morgan et al.  \cite{Morgan03}).   The brightest quasar image
is the  leading image, consistent    with  the arrival time   surfaces
presented in Saha et al. (\cite{Saha06}). Deriving a precise value for
the time delay requires numerical methods.   We use two very different
techniques:   the    minimum    dispersion  method    (Pelt    et  al.
\cite{Pelt96}), and a polynomial fit to the data, e.g., as implemented
in Kochanek et al. (\cite{Kochanek06a}).

Our  full dataset  consists   of 62 observing  epochs.   However,  the
photometric points for    5 of the   epochs, marked  by triangles   in
Fig.~\ref{lightcurve}, deviate very far away from the general trend in
the light curve of the quasar image  B.  These deviating points do not
seem  to   be artifacts  due  to  bad PSF   or  problems  in  the flux
calibration.  However,    they introduce  instable behaviour   of  the
dispersion function and    they do not  reflect  the  otherwise smooth
variation of the quasar.  We choose to remove them from the data,
prior to the time delay measurement. 

\subsection{The minimum dispersion method}

The minimum dispersion method has been  already applied many times
to sparsely sampled light curves of lensed quasars (e.g., Pelt et al.
\cite{Pelt98}). In this method, a guessed value is chosen for the time 
delay.  The light  curve of one quasar  image is taken as a  reference
and the other is  shifted by all  the time delays to  be tested in  an
arbitrarily long interval around the  guess value.  Pairs of points are
formed with each point of  the  reference  curve and its  nearest
corresponding  neighbour in the shifted version   of the second light
curve, and a mean  distance, in magnitude,  is computed between the two
curves. The best value of the time delay is the one that minimises the
{\it dispersion function} constructed in  that way.  The method has to
be   used with  caution, in   particular   when dealing with the  flux
normalisation of the light curves.

The mean magnitude of the data points of both light curves must be set
to  zero prior to  their   cross-correlation.  However, the   required
normalisation factors are  often  determined  from the light    curves
themselves, which can  lead  to a wrong  normalisation as  soon as the
time delay becomes significantly  long compared with the  total length
of the observations.  Ideally,  the normalisation must  be done on the
exact same  portion of the  intrinsic light  curve of  the quasar.  In
lensed quasars, we see several versions of this intrinsic light curve,
which are shifted in magnitude and time, and clipped by the visibility
window.  Determining the factors directly from the data will therefore
lead to a wrong normalisation.

If we normalise  the  data of \obj\ without  taking  into account  the
clipping of the light curves, the dispersion function,  as shown in the
upper left  panel of Fig.~\ref{TDresults1}, reaches  its minimum for a
$\Delta t = 34.4$~days, with a secondary minimum around 55 days.

In order  to estimate the correct normalisation  factors, we follow an
iterative procedure.  We  first estimate a  rough normalisation factor
from  the data and we measure  the minimum D$_{min}$ of the dispersion
function as well as the position $\Delta t$  of this minimum.  We then
slightly change  the magnitude shift $\Delta m$ between the  two light
curves and we  repeat  the (D$_{min}$, $\Delta t$)  measurement,  so
that we can explore the $\Delta  m$ vs.  D$_{min}$ plane.  The  result
is shown in Fig.~\ref{TDresults2}.  
We  consider that  the correct magnitude shift
between  the light curves, and hence,  the best time  delay, is the one
that  minimises  D$_{min}$.   We   obtain  in that  way  $\Delta  t  =
52.6$~days.  Note that  the magnitude  shift that minimises  D$_{min}$
also gives a much cleaner, single-peaked dispersion function, as shown
in the lower left panel of Fig.~\ref{TDresults1}.

The accuracy on the time delay  value is estimated using a statistical
Monte-Carlo method: the data points  in the light curves are  slightly
modified following  a   random Gaussian distribution that  mimics  the
measured photometric  errors, and the algorithm   is run on  these new
modified data.  The operation is  repeated  100'000 times.  The  final
time delay is  the mean of  the 100'000 measurements and the 1$\sigma$
error on this value is the error on this mean value. The result of the
Monte-Carlo  runs     is   shown     in    the   right     panels   in
Fig.~\ref{TDresults1}.  Our final value  of  the time  delay using  57
epochs and the   minimum dispersion method is $\Delta   t =  52.3  \pm
1.6$~days.

Finally, note that these values do not take microlensing into account.
The minimum dispersion method is, however, not very sensitive to 
low amplitude microlensing variations, as shown in  Eigenbrod  et
al.~(\cite{Eigenbrod05}). Adding microlensing does not change the
time delay obtained above, but only slightly enlarges the distribution.

\subsection{Polynomial fit to the light curves}
\label{polfit}

A completely different approach  to measure the time  delay is to  fit
fully analytical functions to  the light curves.  An implementation of
this     type   of   method   has     been    used  in   Kochanek   et
al.~(\cite{Kochanek06a}) and in Morgan et al.~(\cite{Morgan06}), where
Legendre polynomials were simultaneously fitted to the light curves of
each quasar image.  The time delay between each pair  of images is one
of the parameters of the fit, as well as the flux ratio of the images,
corrected by the time delay.  Our own  implementation of the  method does not
use any  stabilisation or smoothing term,   which is not  mandatory as
long  as the polynomial order is  chosen in adequation with the number
of data points  in the light  curves.  In addition, a slow photometric
variation is added to each light curve,  in order to take microlensing
into account.

\begin{figure}[t]
\begin{center}
\leavevmode
\includegraphics[width=4.3cm]{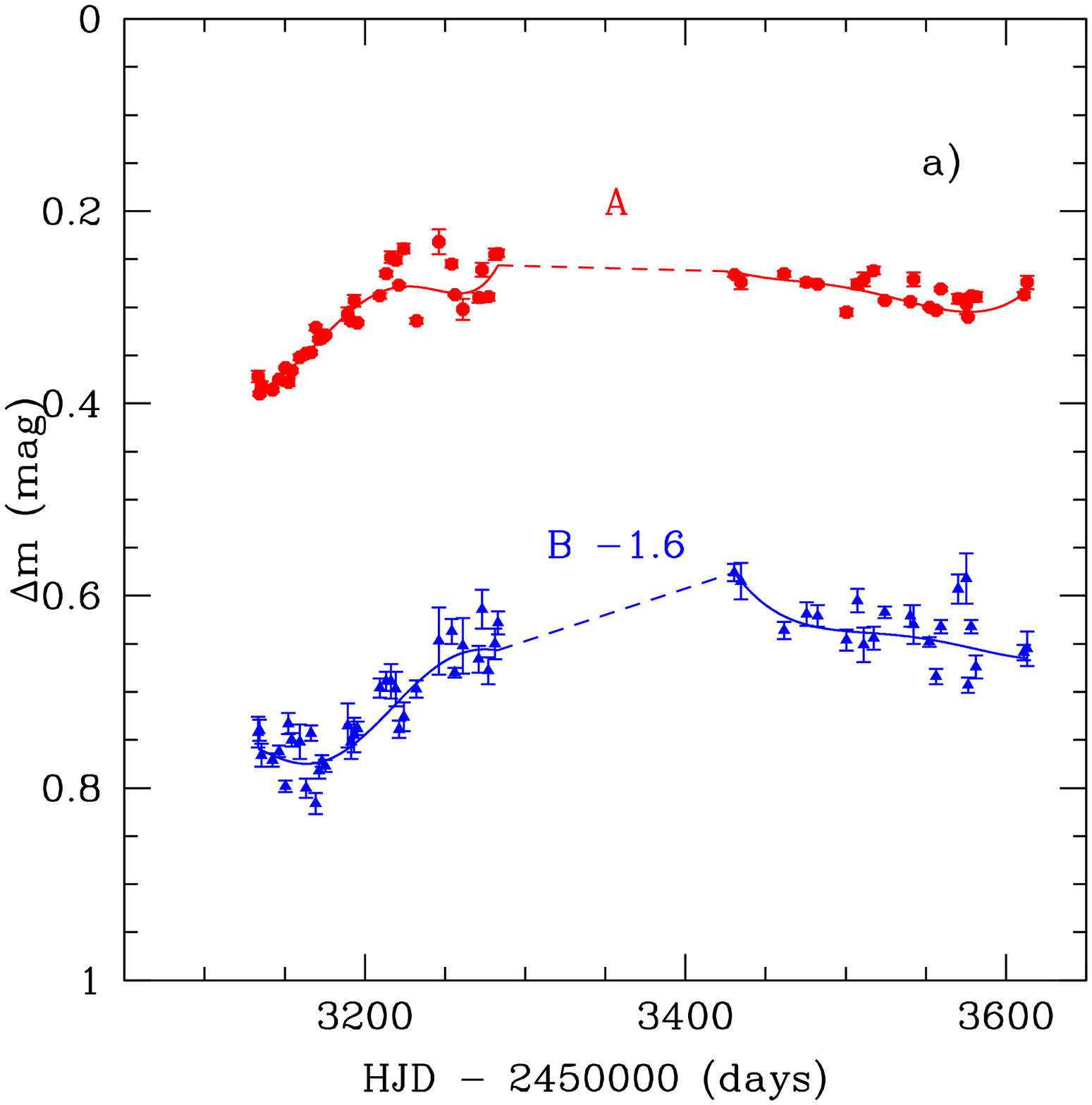}
\includegraphics[width=4.3cm]{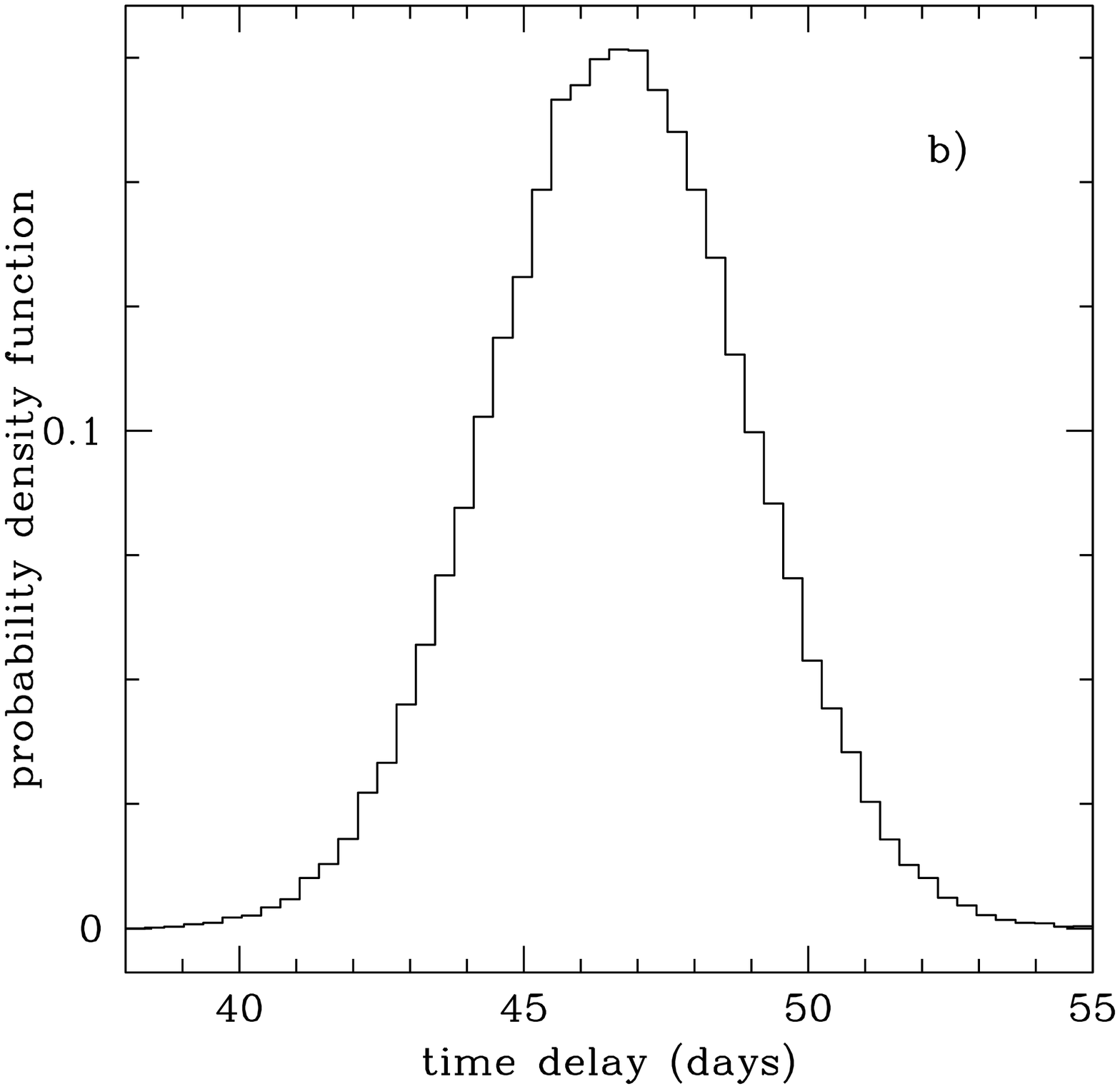}
\includegraphics[width=4.3cm]{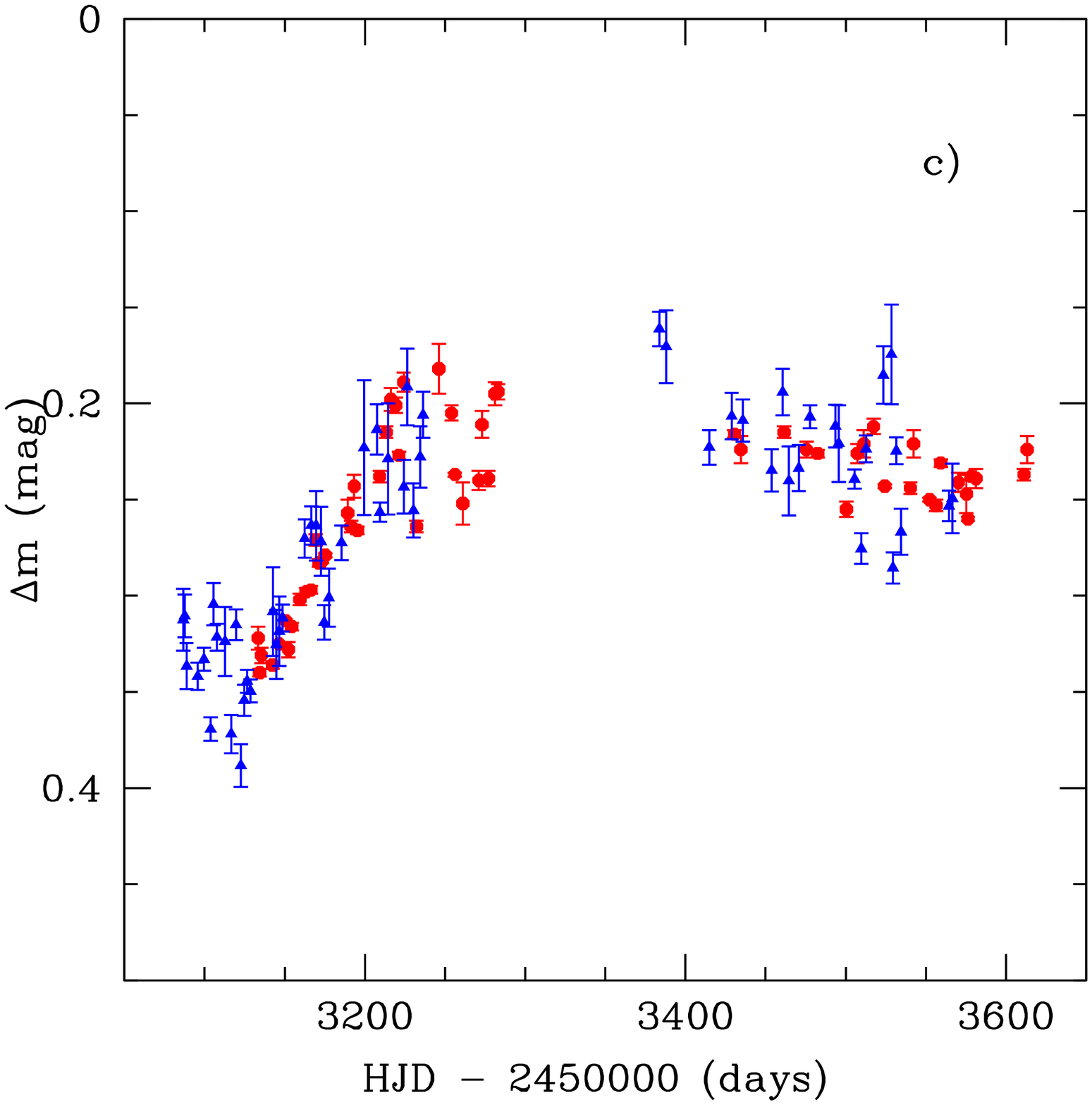}
\includegraphics[width=4.3cm]{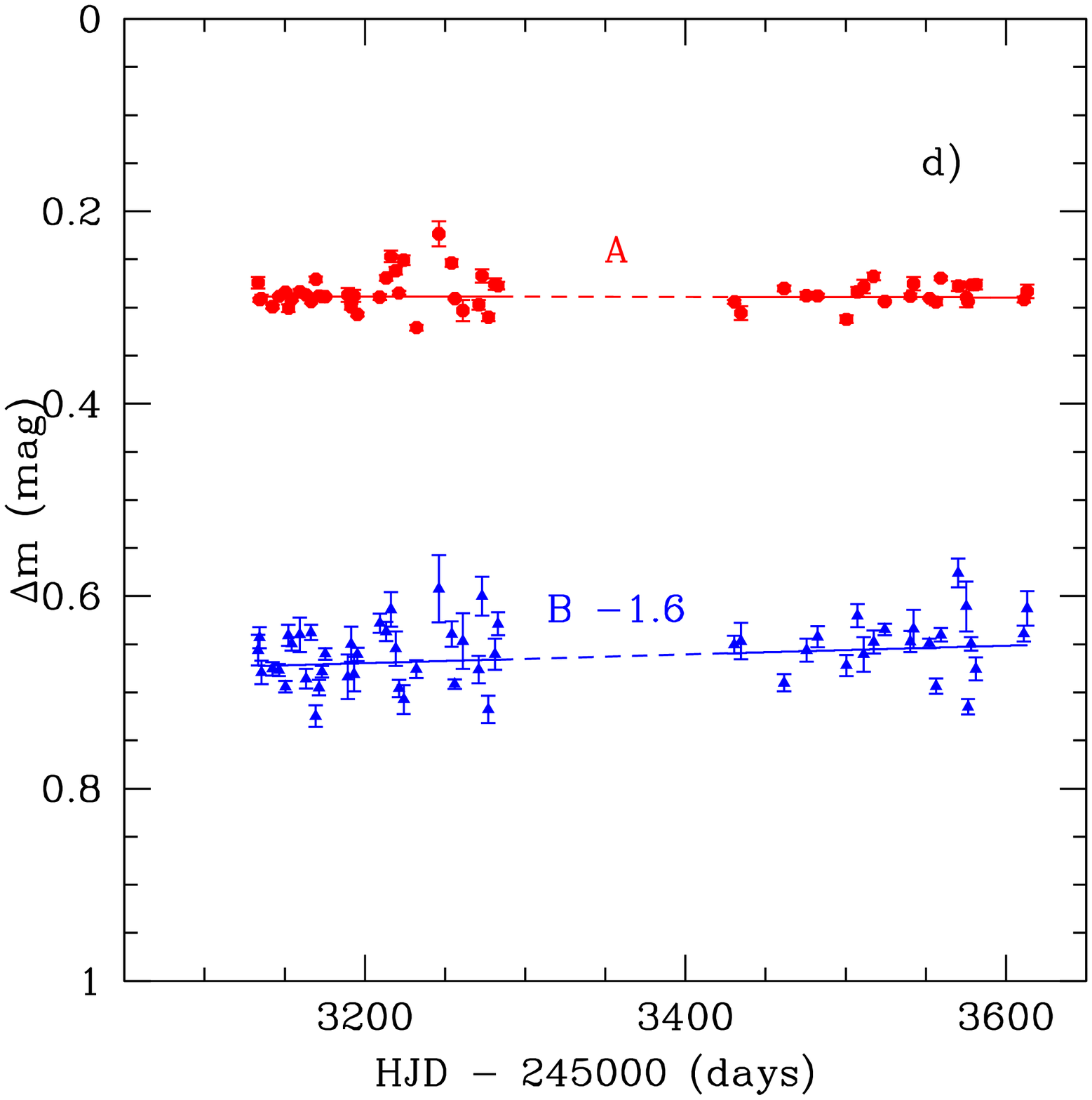}
\caption{a) Best polynomial fit to the light curves. The order of the
  Legendre polynomial is 5 for the first season and 4 for the second
  season. 
b) Result of the  Monte-carlo simulation for 100'000  modified
light  curves, leading to  a mean time delay of  $\Delta t  = 46.7 \pm
2.2$~days.
c) Light curves shifted by the time delay and corrected for
slow microlensing variations.
d) Microlensing variations of quasar image B relative to quasar image
A, taken  as a slow linear trend over the whole period of observation.}
\label{TDevans1}
\end{center}
\end{figure}

This method is applied   to the data   in two  ways, which are   fully
equivalent for   a    double quasar  with  little   contribution   of
microlensing.   First,   we assume that   the  intrinsic   photometric
variation of the quasar is well represented  by the light curve of the
brightest quasar image  A.  The resulting microlensing contribution to
the total variation in the light curve of the quasar image B then only
reflects  differential microlensing  relative to  the  component A.  A
second way to implement  the method, is to use  one single light curve
to represent  the intrinsic  variations of the  quasar and  to fit  it
simultaneously to the two quasar images.  Two independent microlensing
curves  are  fitted to  the two light  curves,  allowing us to recover
absolute microlensing curves.  The fit to the data  is performed in an
iterative  way and the  algorithm  is  run 100'000  times on  modified
versions of the light curves following  the statistical 1$\sigma$ errors
on the photometry.  As for the minimum dispersion  method, the mean of
the time delay distribution obtained in that  way is taken as the time
delay measurement,  and its standard deviation  as the 1$\sigma$ error
(Fig.~\ref{TDevans1}).

A critical step in the use of this fitting method  is the choice of an
optimal  order for the Legendre polynomials.   The data consist in two
observing seasons, for which  peak-to-peak  variations  and lengths  
are different. We therefore fit  polynomials with different orders for the
two seasons.  
We  progressively  increase  the degree   of  the polynomial until the
$\chi^2$ to the fit  does not significantly improve.   Microlensing is
estimated   in the  same way, but   we   assume that the  microlensing
variations    are slow compared to the     intrinsic variations of the
quasar. We  therefore fit a single  polynomial for both seasons.  Fast
microlensing,  acting on  time-scales of   a  week,  is  seen as    an
additional source of noise, and not as  a systematic effect.  Although
we  increase the degrees  of the intrinsic and microlensing variations
separately when  trying    different  fits,  both   contributions  are
simultaneously fitted to the data,  once the orders of the  polynomial
are chosen.   The optimal combination of  polynomial orders is N=5 for
the first season  of observation and  N=4  for the  second season. The
microlensing variations are modeled using a  simple linear slope.  The
difference between the fit  and the data is  shown in the lower  right
panel of Fig.~\ref{TDevans1},  where the dispersion between the points
is well compatible with the error bar on the individual points.

The  two ways  of applying the  fitting  method to the  data are fully
equivalent  in the particular  case of  \obj, as  we only use  a first
order  polynomial to model the microlensing  and as there are only two
light curves available.  The time delays  obtained in the two ways are
indeed in agreement and yield $\Delta t  = 46.7 \pm 2.2$~days. We find
that slow  microlensing  is almost negligible in   \obj, with a global
variation of 0.02  magnitude over 500 days  (see Fig.~\ref{TDevans1}).
Taking  the mean   of the  two  values  obtained   using the   minimum
dispersion  method and using the  analytical fitting gives $\Delta t =
49.5 \pm  1.9$~days, which we  take as our final  estimate of the time
delay. This translates into a relative error of 3.8\,\%, which is also
in  agreement  with  the predicted     error  bar from   Eigenbrod  et
al.~(\cite{Eigenbrod05})   for   light     curves   with    the   same
characteristics as the ones of \obj.  Finally, the $R$-band flux ratio
between the  quasar images, corrected for the  time delay and the slow
microlensing is $F_{\rm A}/F_{\rm  B}  =$   6.2. 
Note that this flux ratio is constant in the time delay range.


\section{Parametric modeling}
\label{models}

The observational constraints available so far  to model the potential
well in \obj\ consist of the relative astrometry  of the lensed images
A and  B  with   respect  to the  lensing  galaxy   G  (Morgan et  al.
\cite{Morgan03}), and  the image   flux ratio measured   in Section~4.
Lacking a  direct spectrum  of the lensing   galaxy, we assume  a lens
redshift  \zl\ = 0.58,  as  deduced from the  MgII  and FeII absorption
lines observed  in the  spectra of  the  lensed quasar (Morgan  et al.
\cite{Morgan03}).    This value of the lens   redshift  drives all the
following estimates of the Hubble constant, which scales as (1+\zl).

Using the \texttt{lensmodel} package (v 1.08; Keeton \cite{KEE01}), we
fit the lens system with (i) a Singular Isothermal Sphere  model plus
external shear (SIS) and (ii) a de Vaucouleurs model (DV).
Due  to  the misalignment between components  A,  B and G  of the lens
system, we break the circular symmetry of the lens potential by adding
external   shear  to   the   models.  While  it    is  likely that the
gravitational  potential of the  lensing  galaxy is not  circular, the
present observational data  available  for \obj\  do  not offer enough
constraints to  allow the simultaneous  fit of an elliptical potential
plus external shear $\gamma$. However, the time delay of doubly imaged
quasars where the lens lies almost on the line joining the two quasar
images depends little on the structure of the quadrupole term of the
potential (Kochanek \cite{KOC02}).  

The constraints on  our  models are  the  relative astrometry  of  the
lensed images with  respect to the  lensing galaxy  (using a 1$\sigma$
error  bar of   0.002\arcsec\ on   the  positions of   A   and B   and
0.03\arcsec\ for G) and the flux ratio between A  and B as found after
correction   from  time  delay  and  observed   relative  microlensing
(i.e.    $F_{\rm A}/F_{\rm  B}  =$   6.2; Sect.~\ref{timedelays}). We
consider  a 5\,\% uncertainty  on the flux   ratio in order to include
differential  dust extinction due  to the lensing galaxy, as suggested
by the multi-color data of Morgan  et al.  (\cite{Morgan03}).  
The modification of the flux ratio  due to the time delay uncertainty or
microlensing is  negligible over the period  of observation, 
as pointed out in Section 4.  In addition, our flux ratio
does  not differ much from that  of Morgan  et al.  (\cite{Morgan03}),
obtained 13  months   before the   start  of  our   observations,  but
uncorrected for the time delay.

\begin{figure}[t]
\begin{center}
\includegraphics{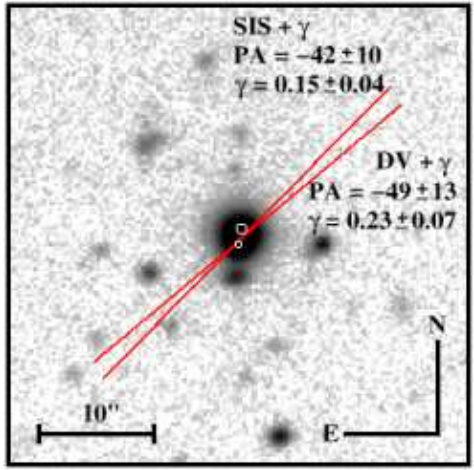}
\caption{A zoom on the $R$-band image of \obj\ obtained at 
Maidanak   Observatory (see Fig.~\ref{field}).    The positions of the
quasar images are marked by white circles.  The direction of the shear
for  our SIS+$\gamma$  and for  the   DV model   is  indicated by  two
lines. It does not point  towards any specific  galaxy in the field of
view. Its  large  amplitude makes  it  mandatory to model  the overall
lensing potential.
}
\label{zoom}
\end{center}
\end{figure}

The  SIS+$\gamma$  model has  just   enough parameters to   be exactly
constrained  by the  data (i.e. zero   degree of freedom),  but the DV
model has  one  more parameter,   the effective  radius $R_e$ of   the
lensing  galaxy,  that is unknown  with the  existing imaging  data of
\obj. We therefore  run our models with  a fixed effective radius.  We
consider a wide range of plausible effective radii, $0.2$~kpc $< R_e <
20$~kpc, and take the two boundaries for our calculations. These values
enclose the range  of measured effective  radii in the Sloan-Lens  ACS
Survey (SLACS)  sample of  15  early-type lens  galaxies (Treu et  al.
\cite{TRE06}).   They correspond to  angular scales of $\theta_e \sim$
0.3\arcsec\ and 3.0\arcsec\ at   the  lens redshift, adopting \ho\   =
70 \kmsmpc, \om\ = 0.3, \ol\ = 0.7.

Using the time delay measured in Section~\ref{timedelays}, $\Delta t =
49.5 \pm 1.9$~days, we  find \ho\ = 51.7$^{+4.0}_{-3.0}$~\kmsmpc\ with
the SIS+$\gamma$ model.   The DV+$\gamma$ model,  with the lower bound
value   of  $\theta_e     =  0.3$\arcsec,    yields    \ho\  =    80.8
$^{+7.0}_{-3.0}$~\kmsmpc.   Using   the  upper  value    $\theta_e   =
3.0$\arcsec, the  DV model gives  \ho\ = 55.1$^{+5.1}_{-3.8}$ \kmsmpc.
Clearly only large values of the effective radius $\theta_e$  (ie
$\theta_e > 3''$) reconcile the value of \ho\ given by the
SIS+$\gamma$ and by the deVaucouleurs models. 
On the other hand, \ho = 72 \kmsmpc\ is found for  $\theta_e =$
0.8\arcsec.

As previously argued, the unkown lens ellipticity is not a major
source of uncertainty on the predicted value of \ho. In order to probe
its effect, we fitted on the observed astrometry and flux ratio, 16200
elliptical lens models with fixed ellipticity $e$, lens major axis
position angle $\phi$ and  shear position angle $\theta_{\gamma}$,
uniformly distributed in the ranges [0:0.4] ($e$) and
[-90\degr:90\degr] ($\phi$ and $\theta_{\gamma}$). Following this
procedure, we find that including the ellipticity in the lens model
does not enlarge the distribution of predicted values of \ho.

All the 1$\sigma$ uncertainties quoted above are deduced by evaluating
\ho\ from 1$\sigma$ $\Delta \chi^2$ contours in the plane
$\gamma$-$\theta_{\gamma}$ (i.e. $\Delta \chi^2 = 2.3$ for two degrees
of freedom).  The  external  shear predicted by both  the SIS+$\gamma$
and   DV  models  is  large:    $\gamma_{\rm  SIS} =   0.15 \pm  0.04$
($\theta_\gamma = -42^{\circ} \pm 10$) and  $\gamma_{\rm DV} = 0.23 \pm
0.07$    ($\theta_\gamma =   -49^{\circ}  \pm  13$),   for $\theta_e =
0.3$\arcsec.  The  direction of  the   shear does not  point   towards
obvious specific external  perturber  in  Fig.~\ref{zoom}.   Its large
amplitude suggests, however, that other mass  clumps along
the line of sight do modify the overall potential well.


\section{Conclusions}

The  main result   of  the  present  work   is the  first  time  delay
measurement of the COSMOGRAIL project, for the doubly imaged quasar
\obj. Our best estimate of the time delay is $\Delta t  = 49.5 \pm 1.9$~days (1$\sigma$). This
corresponds to a relative accuracy of 3.8\,\%. The $R$-band flux ratio
of the two quasar images, corrected for microlensing  and for the time
delay, is $F_{\rm A}/F_{\rm B} = 6.2 \pm 5\%$.

The amplitude of the external shear in  the circular lens models
is larger than $\gamma=0.15$, and suggests that both external shear
and ellipticity in the  main  lensing galaxy are  necessary to  model
the total potential well. However,  the  present observational
constraints  available  for \obj\, prevent   us from introducing  a
lens ellipticity and position angle  in the models.   Deep, high
resolution  images of \obj\ will be necessary to estimate the latter
two parameters. 

With  the present   3.8\,\%   accuracy  on   the time  delay,  we  can
efficiently   discriminate   between  families  of   lens models   or,
conversely, estimate \ho\ assuming a  lens model.  
Our results suggest that SIS models are not acceptable
in order to match the current favored value of the Hubble constant
(\ho\  = 72 $\pm$ 8 \kmsmpc) and that only models of the lensing
galaxy that have a de Vaucouleurs mass profile can reproduce it.

Slow microlensing  is  negligible   in   the first   two  seasons   of
monitoring, with a  global variation  of  less than 0.02 mag  over 500
days. 

Finally, note that the redshift of the lensing galaxy is based on MgII
and  FeII  absorption  lines  in     the   spectrum of  the     quasar
images. Although the impact parameter necessary to form these lines is
small,   we cannot  exclude   that the true   lens  redshift does not
correspond to the absorption lines, as in other systems (HE~1104-1805;
Lidman et  al.~\cite{lidman2000}). A  genuine spectrum of  the lensing
galaxy must be obtained to confirm its redshift.

\begin{acknowledgements}
  COSMOGRAIL  is financially supported   by the Swiss National Science
  Foundation  (SNSF).   Pierre Magain is   financially  supported by
  the  Belgian Science Policy (BELSPO) in  the  framework of the
  PRODEX Experiment Arrangement  C-90195.  M.  Ibrahimov  is
  partially supported by the Swiss National Science Foundation (SNSF)
  and EPFL  in the context of COSMOGRAIL.
\end{acknowledgements}

\begin{table*}[t!]
\caption[]{Photometry of \obj\ and of reference star \#5, as in 
Fig.~\ref{lightcurve}. The Julian date corresponds to HJD-2450000 days. 
The five points marked by an asterisk are not used in the determination
of the time delay.}
\label{data}
\begin{center}
\begin{tabular}{cccccccc}
\hline\hline
HJD & seeing ["] & mag A & $\sigma_A$ & mag B & $\sigma_B$ & mag star \#5 & $\sigma_{star \#5}$ \\
\hline
3133.412 & 1.5 & 0.273 & 0.006 & 2.244 & 0.016 & -0.175 & 0.013 \\
3134.362 & 1.1 & 0.291 & 0.002 & 2.241 & 0.011 & -0.166 & 0.004 \\
3135.385 & 0.9 & 0.282 & 0.004 & 2.268 & 0.012 & -0.177 & 0.008 \\
3142.386 & 1.3 & 0.287 & 0.002 & 2.273 & 0.007 & -0.167 & 0.003 \\
3146.415 & 1.2 & 0.276 & 0.001 & 2.264 & 0.006 & -0.178 & 0.002 \\
3150.406 & 0.9 & 0.264 & 0.001 & 2.300 & 0.006 & -0.176 & 0.001 \\
3152.367 & 1.0 & 0.279 & 0.004 & 2.231 & 0.011 & -0.172 & 0.008 \\
3154.407 & 0.9 & 0.267 & 0.002 & 2.252 & 0.007 & -0.175 & 0.003 \\
3159.364 & 1.1 & 0.253 & 0.003 & 2.254 & 0.018 & -0.174 & 0.005 \\
3163.315 & 1.7 & 0.249 & 0.002 & 2.301 & 0.010 & -0.173 & 0.003 \\
3166.425 & 1.0 & 0.248 & 0.002 & 2.244 & 0.008 & -0.177 & 0.003 \\
3169.393 & 1.1 & 0.222 & 0.003 & 2.317 & 0.011 & -0.189 & 0.006 \\
3171.430 & 1.5 & 0.234 & 0.002 & 2.284 & 0.008 & -0.172 & 0.004 \\
3173.298 & 1.1 & 0.233 & 0.001 & 2.275 & 0.006 & -0.168 & 0.002 \\
3175.300 & 1.1 & 0.230 & 0.001 & 2.279 & 0.006 & -0.178 & 0.001 \\
3189.265 & 1.1 & 0.208 & 0.007 & 2.236 & 0.023 & -0.172 & 0.012 \\
3191.360 & 1.1 & 0.215 & 0.003 & 2.255 & 0.018 & -0.167 & 0.004 \\
3193.304 & 1.3 & 0.194 & 0.006 & 2.246 & 0.018 & -0.166 & 0.012 \\
3195.252 & 1.0 & 0.217 & 0.002 & 2.239 & 0.007 & -0.170 & 0.004 \\
3203.314* & 1.4 & 0.186 & 0.001 & 2.345 & 0.007 & -0.173 & 0.002 \\
3209.256 & 1.0 & 0.189 & 0.003 & 2.198 & 0.010 & -0.167 & 0.007 \\
3213.256 & 1.0 & 0.166 & 0.003 & 2.190 & 0.010 & -0.183 & 0.007 \\
3216.310 & 1.7 & 0.149 & 0.006 & 2.189 & 0.018 & -0.163 & 0.011 \\
3219.281 & 1.6 & 0.151 & 0.004 & 2.198 & 0.018 & -0.155 & 0.008 \\
3221.234 & 1.0 & 0.178 & 0.002 & 2.241 & 0.009 & -0.179 & 0.003 \\
3224.209 & 1.2 & 0.140 & 0.005 & 2.228 & 0.015 & -0.166 & 0.010 \\
3232.187 & 1.1 & 0.215 & 0.003 & 2.198 & 0.009 & -0.168 & 0.005 \\
3246.163 & 1.3 & 0.133 & 0.013 & 2.148 & 0.035 & -0.142 & 0.026 \\
3248.138* & 1.5 & 0.172 & 0.004 & 2.054 & 0.013 & -0.114 & 0.007 \\
3254.125 & 0.9 & 0.156 & 0.004 & 2.139 & 0.013 & -0.160 & 0.007 \\
3256.135 & 1.0 & 0.188 & 0.001 & 2.181 & 0.005 & -0.173 & 0.003 \\
3261.115 & 1.5 & 0.203 & 0.011 & 2.153 & 0.029 & -0.178 & 0.017 \\
3271.109 & 1.3 & 0.191 & 0.005 & 2.167 & 0.014 & -0.183 & 0.009 \\
3273.103 & 1.2 & 0.163 & 0.007 & 2.117 & 0.020 & -0.176 & 0.013 \\
3277.104 & 1.0 & 0.190 & 0.004 & 2.179 & 0.014 & -0.191 & 0.007 \\
3281.106 & 1.4 & 0.146 & 0.006 & 2.151 & 0.016 & -0.167 & 0.011 \\
3283.100 & 1.2 & 0.146 & 0.004 & 2.129 & 0.012 & -0.164 & 0.008 \\
3430.540 & 0.9 & 0.167 & 0.002 & 2.078 & 0.009 & -0.135 & 0.003 \\
3434.546 & 0.8 & 0.175 & 0.007 & 2.088 & 0.019 & -0.143 & 0.009 \\
3461.499 & 0.9 & 0.166 & 0.003 & 2.138 & 0.009 & -0.150 & 0.005 \\
3475.491 & 0.9 & 0.175 & 0.004 & 2.120 & 0.012 & -0.179 & 0.008 \\
3482.443 & 1.1 & 0.177 & 0.002 & 2.124 & 0.011 & -0.149 & 0.004 \\
3500.415 & 1.3 & 0.206 & 0.004 & 2.147 & 0.011 & -0.168 & 0.008 \\
3507.348 & 1.0 & 0.177 & 0.005 & 2.106 & 0.012 & -0.153 & 0.009 \\
3508.403* & 0.9 & 0.199 & 0.003 & 2.222 & 0.009 & -0.171 & 0.006 \\
3511.303 & 1.0 & 0.172 & 0.007 & 2.153 & 0.018 & -0.151 & 0.012 \\
3517.383 & 0.8 & 0.163 & 0.004 & 2.146 & 0.012 & -0.168 & 0.007 \\
3524.391 & 0.9 & 0.194 & 0.002 & 2.118 & 0.006 & -0.183 & 0.003 \\
3533.412* & 1.6 & 0.193 & 0.002 & 2.002 & 0.007 & -0.171 & 0.004 \\
3540.345 & 1.0 & 0.194 & 0.003 & 2.124 & 0.011 & -0.168 & 0.006 \\
3542.323 & 1.1 & 0.172 & 0.007 & 2.132 & 0.020 & -0.194 & 0.014 \\
3552.291 & 0.9 & 0.201 & 0.001 & 2.149 & 0.005 & -0.179 & 0.001 \\
3556.295 & 0.9 & 0.204 & 0.003 & 2.185 & 0.008 & -0.176 & 0.005 \\
3559.280 & 1.0 & 0.181 & 0.002 & 2.134 & 0.007 & -0.160 & 0.004 \\
3564.295* & 1.3 & 0.228 & 0.001 & 2.300 & 0.007 & -0.167 & 0.003 \\
3570.247 & 1.0 & 0.192 & 0.005 & 2.094 & 0.015 & -0.165 & 0.010 \\
3575.323 & 1.3 & 0.198 & 0.010 & 2.084 & 0.026 & -0.172 & 0.019 \\
3576.264 & 1.0 & 0.211 & 0.001 & 2.195 & 0.008 & -0.181 & 0.002 \\
3578.275 & 1.2 & 0.189 & 0.002 & 2.133 & 0.007 & -0.177 & 0.003 \\
3581.284 & 1.2 & 0.190 & 0.005 & 2.176 & 0.012 & -0.188 & 0.009 \\
3611.225 & 1.2 & 0.188 & 0.003 & 2.161 & 0.008 & -0.165 & 0.005 \\
3613.201 & 1.5 & 0.175 & 0.007 & 2.156 & 0.018 & -0.180 & 0.013 \\
\hline
\end{tabular}
\end{center}
\end{table*}

\end{document}